\documentclass[aps,prl,twocolumn,superscriptaddress,showpacs]{revtex4}
\usepackage{graphicx,bm}

\begin{document}
\title{Mott gap excitations in twin-free YBa$_2$Cu$_3$O$_{7-\delta}$
($T_{\rm c}$ = 93 K) studied by RIXS}
\author{K. Ishii}
\email{kenji@spring8.or.jp}
\affiliation{Synchrotron Radiation Research Center, Japan Atomic
Energy Research Institute, Hyogo 679-5148, Japan}
\author{K. Tsutsui}
\affiliation{Institute for Materials Research, Tohoku University,
Sendai 980-8577, Japan}
\author{Y. Endoh}
\affiliation{International Institute for Advanced Studies,
Kizugawadai, Kizu, Kyoto 619-0025, Japan}
\author{T. Tohyama}
\affiliation{Institute for Materials Research, Tohoku University,
Sendai 980-8577, Japan}
\author{K. Kuzushita}
\affiliation{Synchrotron Radiation Research Center, Japan Atomic
Energy Research Institute, Hyogo 679-5148, Japan}
\author{T. Inami}
\affiliation{Synchrotron Radiation Research Center, Japan Atomic
Energy Research Institute, Hyogo 679-5148, Japan}
\author{K. Ohwada}
\affiliation{Synchrotron Radiation Research Center, Japan Atomic
Energy Research Institute, Hyogo 679-5148, Japan}
\author{S. Maekawa}
\affiliation{Institute for Materials Research, Tohoku University,
Sendai 980-8577, Japan}
\author{T. Masui}
\affiliation{Superconducting Research Laboratory, ISTEC,
Tokyo 135-0062, Japan}
\author{S. Tajima}
\affiliation{Superconducting Research Laboratory, ISTEC,
Tokyo 135-0062, Japan}
\author{Y. Murakami}
\affiliation{Synchrotron Radiation Research Center, Japan Atomic
Energy Research Institute, Hyogo 679-5148, Japan}
\affiliation{Department of Physics, Tohoku University, Sendai
980-8578, Japan}
\author{J. Mizuki}
\affiliation{Synchrotron Radiation Research Center, Japan Atomic
Energy Research Institute, Hyogo 679-5148, Japan}
\date{\today}

\begin{abstract}
Mott gap excitations in the high-$T_{\rm c}$ superconductor of the
optimal doped YBa$_2$Cu$_3$O$_{7-\delta}$ ($T_{\rm c}$ = 93 K) have
been studied by the resonant inelastic x-ray scattering
method. Anisotropic spectra in the $ab$-plane are observed in a
twin-free crystal. The excitation from the one-dimensional CuO chain
is enhanced at 2 eV near the zone boundary of the $\bm{b}^{\ast}$
direction, while the excitation from the CuO$_2$ plane is broad at
1.5-4 eV and almost independent of the momentum transfer. Theoretical
calculation based on the one-dimensional and two-dimensional Hubbard
model reproduces the observed spectra by taking the different
parameters of the on-site Coulomb energy. The fact of the Mott gap of
the CuO chain site is much smaller than that of CuO$_2$ plane site is
observed for the first time.
\end{abstract}


\maketitle

Among various superconducting copper oxides
YBa$_2$Cu$_3$O$_{7-\delta}$ (YBCO) is still recognized to be an
important class of materials in order to elucidate the mechanism of
the unconventional superconductivity with high transition
temperature. However, the unique crystal structure including the
one-dimensional (1D) CuO chain running along the crystalline $\bm{b}$
axis next to the double CuO$_2$ planes gives rise to a complexity.
The CuO chain contributes to the bulk electronic properties such as
the optical conductivity \cite{Koch1} and the dc electric conductivity
\cite{Takenaka1}. Furthermore a substantial anisotropic character in
the superconducting state observed in the far infrared spectra
\cite{Basov1} and the thermal conductivity \cite{Gagnon1} suggests
that the superfluid density is induced in the CuO chain, though it is
generally agreed that the two-dimensional (2D) CuO$_2$ planes are the
most important element for the superconductivity.  An issue remains
how the CuO chain plays a role in the superconductivity of YBCO. In
other words, the instability of the 1D metals of CuO chain itself
might appear as a charge modulation of the twice of the Fermi wave
vector ($2k_F$) \cite{Edwards1,Derro1,Maki1}, but there might be more
than the simple proximity effect of the superconductivity in the
CuO$_2$ planes. These experimental evidences have been discussed
either by the proximity effect induced chain superconductivity of the
CuO$_2$ plane with the magnetic impurities \cite{Morr1} or the Friedel
oscillation in the fragmentary CuO chain due to the partial oxygen
depletion \cite{Mori1}. Therefore any experimental evidences to
clarify the role of the CuO chain would make a significant
contribution to understand the general consequence of high transition
temperature in YBCO.

As a first step to solve this issue, the intrinsic electronic
structure of the chain should be clarified. One reason why the
electronic structure remains unclear is the lack of experimental tool
to differentiate the electronic states of the chain from that of the
plane. Here we applied the resonant inelastic x-ray scattering (RIXS)
method by which the momentum dependence of the electronic excitation
can be measured in contrast to the conventional optical method. The
angle-resolved photoemission spectroscopy (ARPES), which also gives
momentum-resolved spectra, presented the dispersion relation below
$E_F$ in the CuO chain \cite{Lu1}, but the excitations across $E_F$
have not been searched yet. The ARPES essentially yields the one
particle spectra for the occupied states below the Fermi energy
($E_F$), while the RIXS gives the two particle excitations. The RIXS
results of the Mott insulators, such as cuprates
\cite{Hasan1,Hasan2,Kim1,Kim2,Kim3} and a manganite \cite{Inami2},
show a peak structure in the excitation spectra crossing so called
Mott gap. Furthermore, recent RIXS measurements of hole-doped
manganites showed that a salient peak feature of the Mott gap remains
even in the metallic state and the energy gap is partly filled at the
same time \cite{Ishii1}, which really demonstrates the capability of
RIXS to explore the electronic excitations in the metallic state of
the transition metal oxides.

In this letter, we present the RIXS spectra of the twin-free YBCO of
the optimal doping ($T_{\rm c}$ = 93 K), from which we can
successfully distinguish the electronic excitations between the
CuO$_2$ planes and the CuO chain. We also show the anisotropic x-ray
absorption spectra near the Cu $K$-edge. It is noted that we selected
the optimally doped YBCO because the anisotropy in both the normal
conductivity and the superconductivity are saturated.

The experiments were carried out at the beam line 11XU of SPring-8
\cite{Inami1}. A Si (111) double-crystal monochromator and a Si (400)
channel-cut secondary monochromator were utilized. Horizontally
scattered x-rays were analyzed in energy by a spherically bent Ge
(733) crystal. Overall energy resolution is about 400 meV estimated
from the full width half maximum (FWHM) of the quasielastic
scattering. The $\bm{c}$-axis of the crystal was kept perpendicular to
the scattering plane. Our experimental condition of the
$\pi$-polarization of the incident x-ray enables us to reduce the
intensity of the elastic scattering when the scattering angle
($2\theta$) is close to 90 degree.  It is crucially important to
measure the low energy excitation without being disturbed by the tail
of the elastic scattering, so that the Brillouin zone measured is
chosen to be $2\theta$ close to 90 degree. We used two single
crystals. One is twinned crystal which is used for the incident energy
($E_i$) dependence, while x-ray absorption spectra and the momentum
dependence of RIXS were measured for a twin-free crystal. The
twin-free single crystal was grown by a crystal pulling technique
\cite{Yamada1} and detwinned under uniaxial pressure. All the spectra
were collected at room temperature.

\begin{figure}
\includegraphics[scale=0.4]{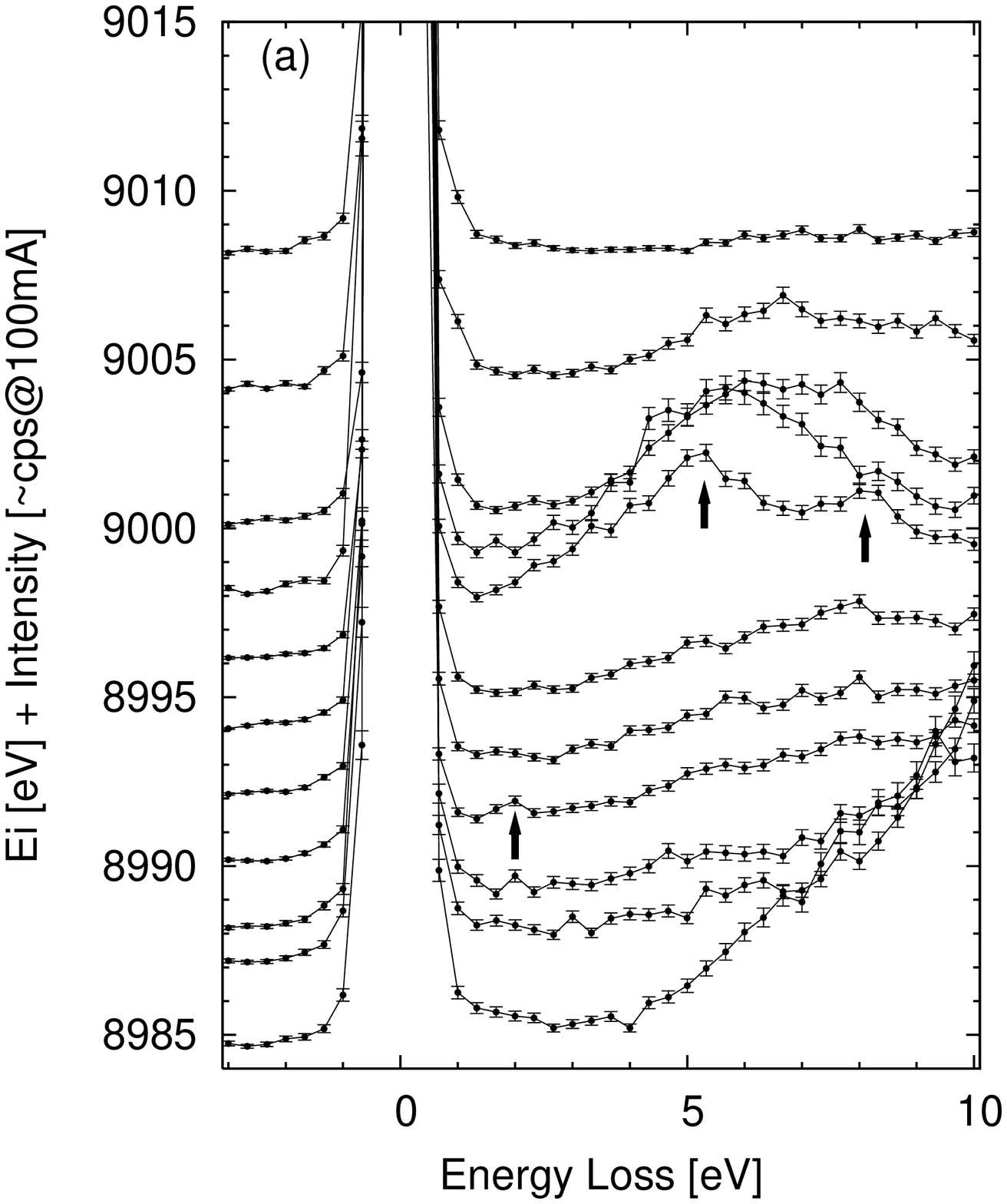}
\includegraphics[scale=0.4]{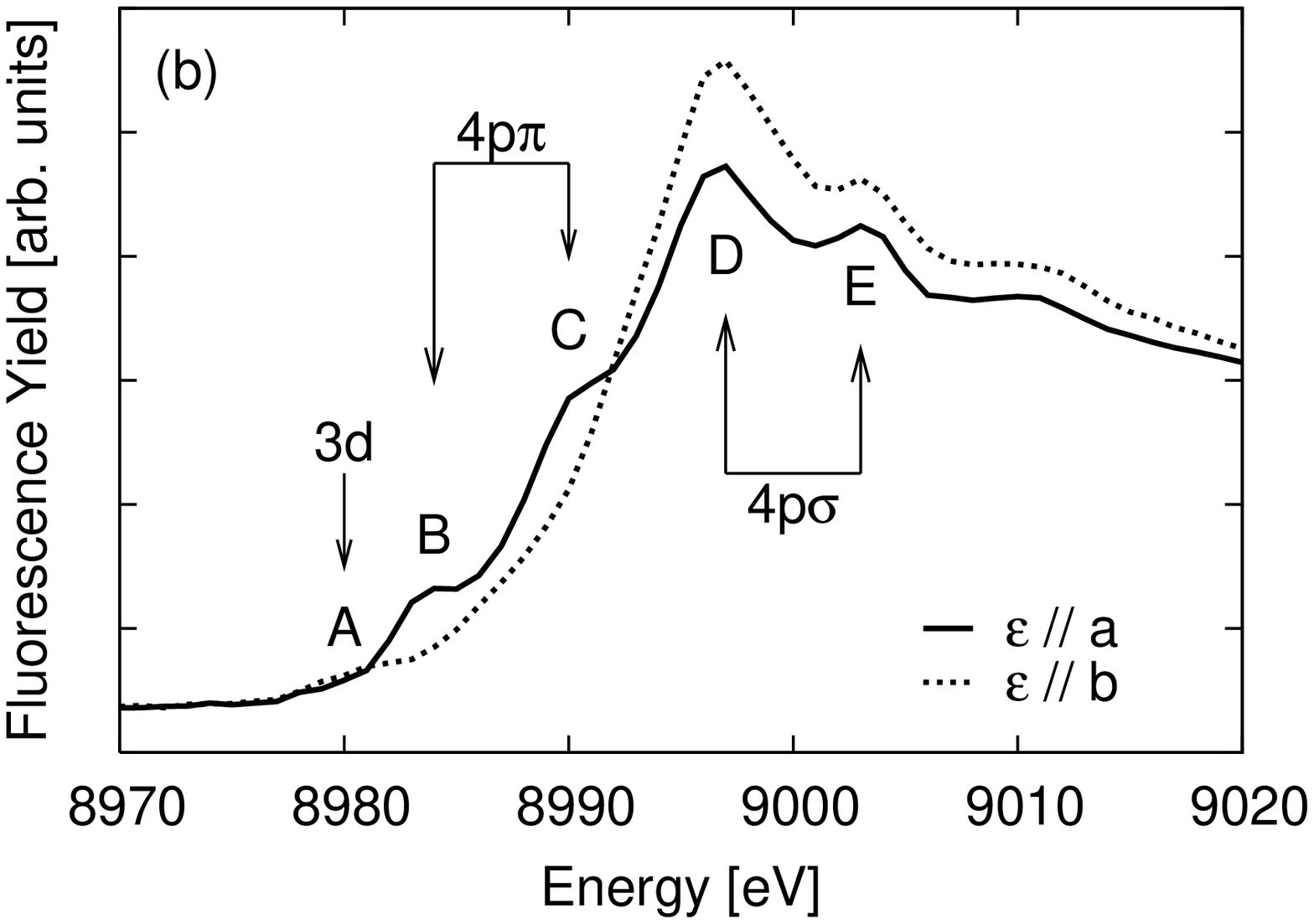}
\caption{(a) Resonant inelastic x-ray scattering spectra of
YBa$_2$Cu$_3$O$_{7-\delta}$ as a function of energy loss at some
representative incident x-ray energies ($E_i$). The scattering vector
is fixed at $\bm{Q}=(4.5,0,0)$. Three resonantly enhanced excitations
are indicated by the arrows. The strong intensity above 10 eV in the
spectrum of $E_i$ = 8984.5 eV comes from the Cu $K\beta_5$ emission
line. (b) Absorption spectra of twin-free YBa$_2$Cu$_3$O$_{7-\delta}$
near the Cu-$K$ absorption edge.}
\label{fig:eidep}
\end{figure}

First we measured the spectra varying the energy of incident x-ray to
determine a resonant energy, as shown in Fig.\ \ref{fig:eidep} (a).
The scattering vector is fixed at $\bm{Q}=(4.5,0,0)$ corresponding to
$\bm{q}=(\pi,0)$ or $(0,\pi)$, where $\bm{q}$ represents the reduced
wave vector in the $ab$-plane. A peak at 2 eV was observed at $E_i$ =
8990 eV. There are another resonant features between 4 and 9 eV at
higher $E_i$; We can see two peaks at 5.5 eV and 8 eV in the spectra
of $E_i$ = 8996 eV.  Hereafter we fixed $E_i$ at 8990 eV to focus on
the excitation at low energy ($\sim 2$ eV).

In Fig.\ \ref{fig:eidep}(b), we show the x-ray absorption spectra of
the twin-free crystal near the Cu-$K$ absorption edge. The spectra
were measured by the fluorescence method. Clear difference between
$\bm{\epsilon}\parallel\bm{a}$ and $\bm{\epsilon}\parallel\bm{b}$ was
observed.  $\bm{\epsilon}$ is the polarization vector of the
x-ray. From the analogy of the spectra of La$_{2-x}$Sr$_x$CuO$_4$ and
Nd$_{2-x}$Ce$_x$CuO$_4$ \cite{Kosugi1}, we can assign the peaks as
follows. The peak at 8980 eV (labeled A in Fig.\ \ref{fig:eidep}(b))
originates from the electric quadrupole transition from $1s$ to $3d$
states. A pair of peaks at 8984 and 8990 eV (B and C) is assigned as
the dipole transition from $1s$ to $4p_{\pi}$ along which there is no
ligand oxygen. The final state of the peak B is well-screened and that
of peak C is poorly-screened. The peaks at 8997 and 9003 eV (D and E)
correspond to the transition to $4p_{\sigma}$ whose orbital extended
toward the ligand oxygens. In the case of
$\bm{\epsilon}\parallel\bm{b}$, the Cu atoms in both the CuO$_2$ plane
and the CuO chain have ligand oxygen along the direction of
$\bm{\epsilon}$, and only the transition to $4p_{\sigma}$ is observed.
On the other hand, because the Cu atoms {\it in the chain} have no
oxygen along the $\bm{a}$-axis, the transition to the $4p_{\pi}$ state
appears in $\bm{\epsilon}\parallel\bm{a}$. The fact that the peak at 2
eV appears resonantly at the intermediate state of $4p_{\pi}$
indicates that the peak originates from the chain.  This inference is
confirmed from the momentum dependence later.

\begin{figure*}
\includegraphics[scale=0.3, trim=0 -5 0 0, clip]{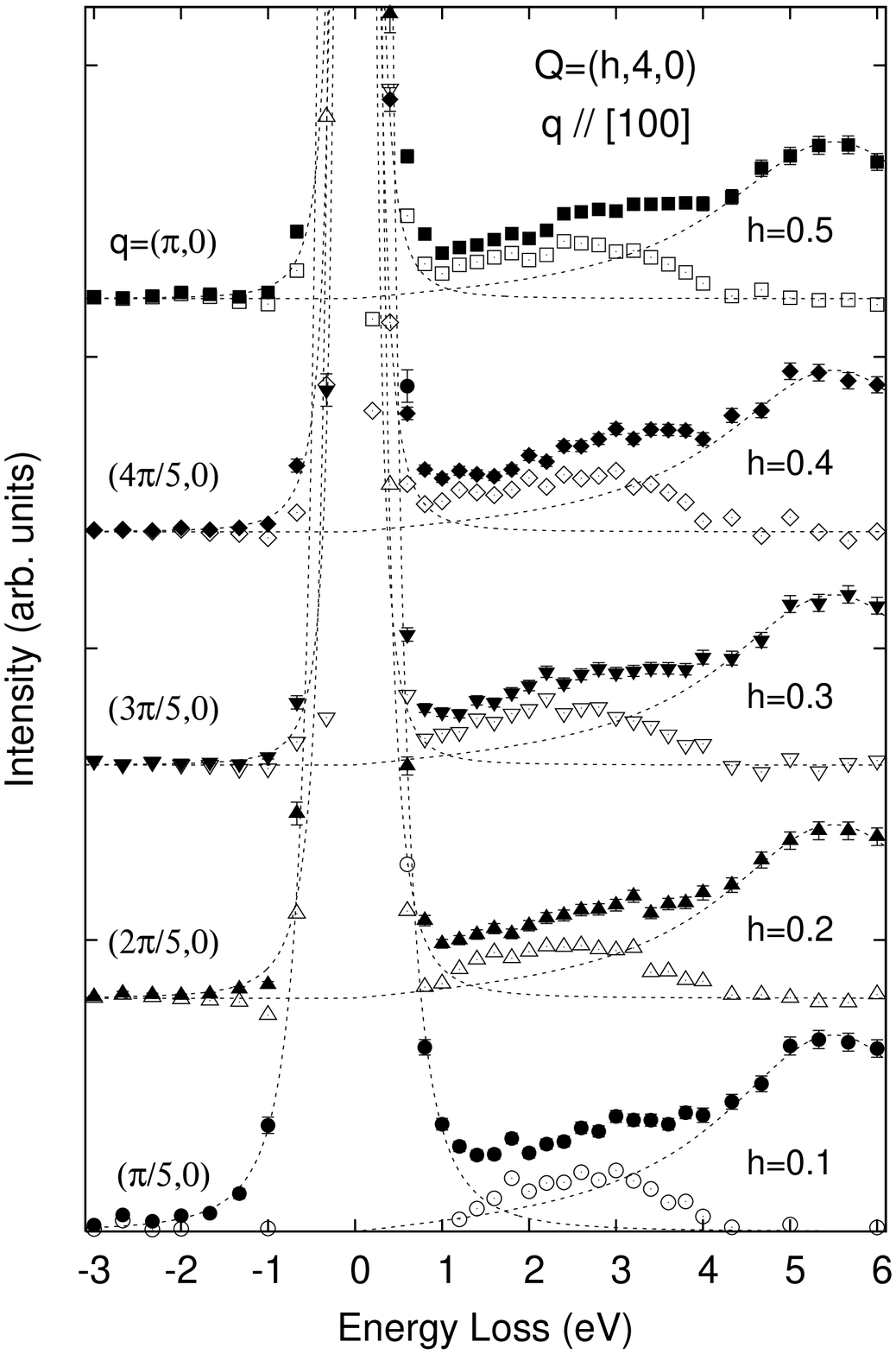}
\includegraphics[scale=0.3, trim=50 -5 0 0, clip]{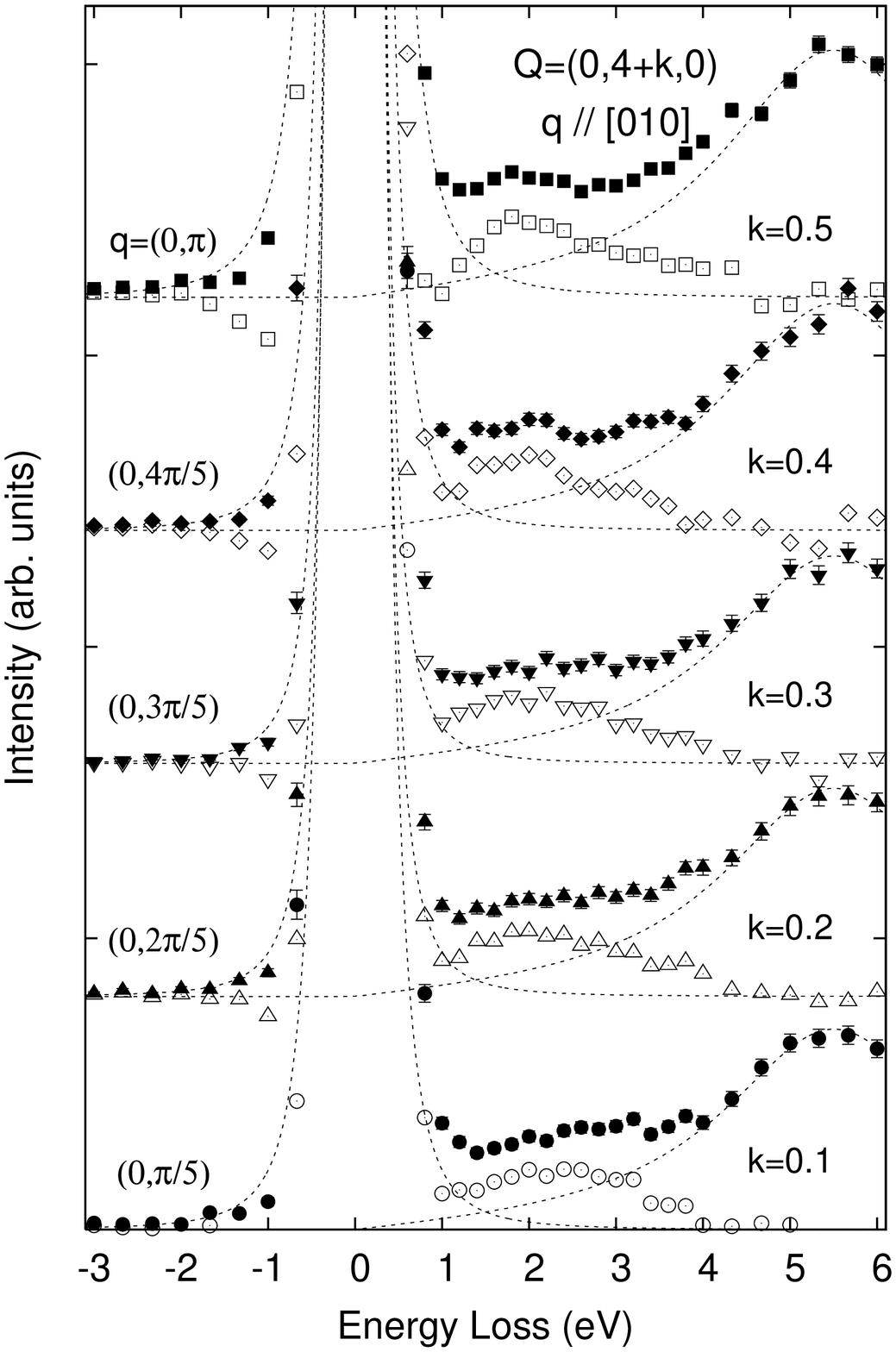}
\includegraphics[scale=0.3, trim=50 -5 0 0, clip]{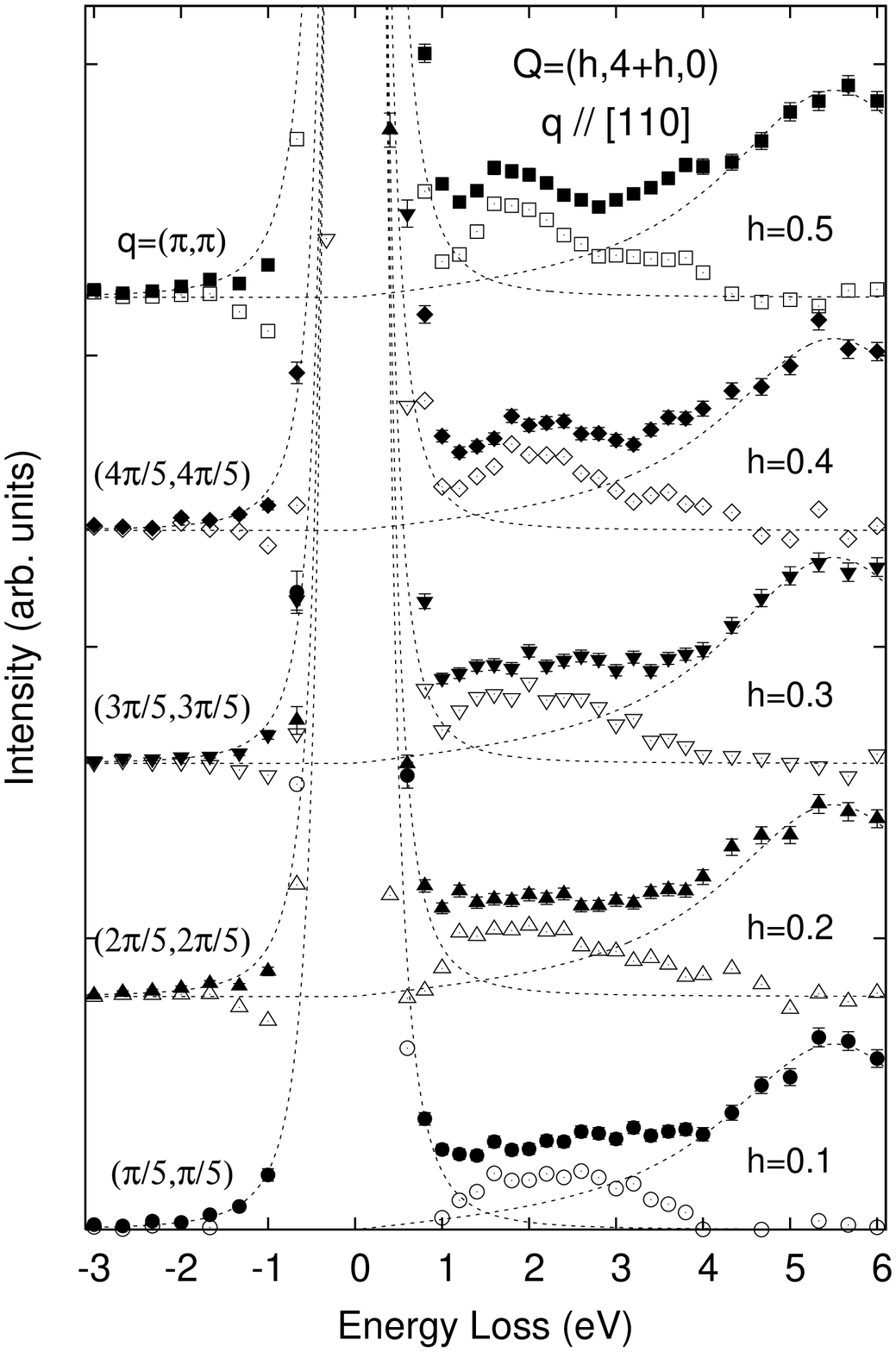}
\caption{Momentum dependence of the RIXS spectra. The energy of
incident x-ray is 8990 eV. The filled marks are the raw data, and the
open marks are the data from which the elastic scattering and the peak
at 5.5 eV indicated by dotted lines are subtracted.}
\label{fig:qdep}
\end{figure*}

Figures \ref{fig:qdep} show the momentum dependence of the RIXS
spectra, in which $\bm{q}$ is parallel to $[100]$, $[010]$, and
$[110]$. The absolute momentum transfer ($\bm{Q}$) is represented as
$\bm{Q}=\bm{G}+\bm{q}/2\pi$ of $\bm{G}=(0,4,0)$.  Apart from the peak
at 5.5 eV, we found two characteristics in the low energy region which
are considered the excitations across the Mott gap. One is a broad
excitation at 1.5-4 eV which is almost independent of the momentum
transfer. The other is an excitation at 2 eV which is prominent at the
zone boundary of $\bm{b}^{\ast}$ direction, that is, the intensity is
enhanced near $(0,\pi)$ and $(\pi,\pi)$. We confirmed that these
characteristics are independent of the selection of the Brillouin
zone. The excitation along the $\bm{a}^{\ast}$- and
$\bm{b}^{\ast}$-axes should be equivalent in the plane. On the other
hand, the momentum dependence along the $\bm{b}^{\ast}$-axis can be
larger than that along the $\bm{a}^{\ast}$-axis in the chain, because
the chain runs along the $\bm{b}$-axis. Accordingly, clear momentum
dependence along the $\bm{b}^{\ast}$-axis of 2 eV peak is a direct
evidence that this is the excitation across the Mott gap in the chain,
while the broad feature at 1.5-4 eV is the excitation in the plane.

\begin{figure}
\includegraphics[scale=0.35]{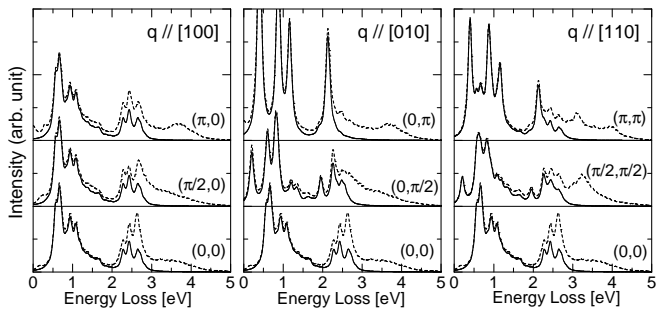}
\caption{Theoretical result of the momentum dependence of the chain
component of the RIXS presented by solid lines. The left, middle, and
right panels show the spectra along [100], [010], and [110],
respectively. The dotted lines denote the spectra which are summed up
with the chain and the plane components, where the plane component is
the same as shown in Ref. \cite{Tsutsui1}.  The model parameters for
the chain are $U/t=4$, $U_c/t=6$. and $\Gamma/t=1$ with $t=0.3$ eV.
The $\delta$-functions are convoluted with Lorentzian broadening of
$0.2t$.  }
\label{fig:theory}
\end{figure}

In order to consider the origin of the peak structures more
accurately, we carried out the calculation of the RIXS spectrum by
using the numerically exact diagonalization technique on small
clusters \cite{Tsutsui1}. We consider the one-dimensional Hubbard
model for the CuO$_3$ chain, where the Zhang-Rice band (ZRB) is
regarded as the lower Hubbard band (LHB). The model includes the
hopping term of the electrons ($t$) and the on-site Coulomb
interaction term ($U$). Because the band calculations and the
angle-resolved photoemission indicate that the filling of the
electrons on the chain is close to one quarter \cite{Pickett1,Lu1}, we
adopt the filling in the twelve-site Hubbard chain. The RIXS spectrum
is expressed as the second-order process of the dipole-transition
between Cu $1s$ and $4p$ orbitals, where the Coulomb interaction
($U_c$) between Cu $1s$ core-hole and the '$3d$-system' is explicitly
included in the process. The values of the model parameters are set to
be $U/t=4$, $U_c/t=6$, and $\Gamma/t=1$ with $t=0.3$ eV where $\Gamma$
is the inverse of the life time of the intermediate state, in order
for the theoretical data to reproduce the 2 eV structure in Fig.\
\ref{fig:qdep}. In order to choose the incident energy for the RIXS
calculation, we first calculated the x-ray absorption spectrum (not
shown here) expressed by the Eq.\ 3 in Ref.\ \cite{Tsutsui2}, and
found the three-peak structure as similar to that of the hole-doped
two-dimensional case \cite{Tsutsui2}: The peaks of the well-screened
and the poorly-screened states are located at around $\omega=U-2U_c$
and $-U_c$, respectively, and the peak that the core-hole is created
at the hole-doped sites is located at around 0. The incident energy
of 8990 eV in the experiment is suggested to be set to the energy of
the poorly-screened state, and thus the incident energy for the
calculation is chosen to the energy of the poorly-screened state.

Figure \ref{fig:theory} shows the theoretical result of the RIXS for
the chain (solid lines), together with the spectrum which is summed up
with the chain and the plane components (dotted lines). Here the
spectrum for the plane is the same as shown in Ref.\ \cite{Tsutsui3}.
The spectrum with zero momentum transfer of the chain site is plotted
for all the momenta along [100] (the left panel in Fig.\
\ref{fig:theory}). The spectrum along [110] is the average of spectra
along [100] and [010]. The spectrum below the energy $\Delta\omega\sim
4t=1.2$ eV ($4t$ is the band width of the free electron) is due to the
excitation within LHB, which is swallowed up by the quasielastic
peak. Note that its spectral weight is extremely reduced when the
incident energy is set to that of the well-screened state and this
feature may be observed by handling with much better energy resolution
in future. The spectrum above $\sim 4t$ is due to the excitation from
the LHB to the upper Hubbard band (UHB). At $(0,0)$, it spreads over
the energy region between 2 and 3 eV. The spectrum becomes more
intensive with increasing the momentum transfer of the chain state,
and the spectral weight concentrates in the narrow energy region at
$(0,\pi)$. The feature is similar to that in undoped chain case
\cite{Tsutsui3}. We consider that this intensive feature for the chain
component appears as the peak structure of 2 eV in the experimental
data.

As shown by the dotted lines in Fig.\ \ref{fig:theory}, there appear
broad spectra in the energy region up to the 4 eV, which originate
from the Mott gap excitation in the plane \cite{Tsutsui2}. The
broad-peak structures correspond to the broad excitations at 1.5-4 eV
seen in Fig.\ \ref{fig:qdep}. The broad feature of the Mott gap
excitations in hole-doped CuO$_2$ plane in YBa$_2$Cu$_3$O$_{7-\delta}$
is quite contrast to that of the undoped one in Ca$_2$CuO$_2$Cl$_2$
\cite{Hasan1} and La$_2$CuO$_4$ \cite{Kim1}, in which a sharp peak can
be seen at 2-4 eV. We can estimate the magnitude of the dispersion in
the plane from the spectra along the [100] direction in
Fig.\ \ref{fig:qdep} because of no dispersion in the chain along this
direction, and it is much smaller than that in undoped ones. The
smaller dispersion in hole-doped CuO$_2$ plane has been predicted in
the theoretical calculation \cite{Tsutsui2}.

Our experimental results and theoretical calculations demonstrate that
the charge gap of the chain is smaller than that of the plane. In
order to compare the theoretical peak position at $(0,\pi)$ with that
of the experimental data, we have to set the value of $U$ to $4t$ for
the chain rather than the value of $10t$ for the plane. The small
value of $U$ suggests that the charge transfer parameter $\Delta$ of
the chain is small compared with that of the plane. The reason of the
different values of $\Delta$'s is because the Cu atom in the plane has
five ligand oxygen while the Cu in the chain has four, so that the
electric environment around the Cu and O sites of the chain is
different from that of the plane \cite{Ohta1}. Optical measurements
for insulating cuprates show that the charge transfer gap decreases
with decreasing the number of ligand oxygen \cite{Tokura1}, which is
consistent with the present case. We also note that the small $\Delta$
for the chain is consistent with a recent study \cite{Mori1}, where it
is shown that the $2k_F$ Friedel oscillation of the charge density,
instead of $4k_F$ one, dominates with small $\Delta$ and it is
discussed that the oscillation can offer the explanation of STM
results \cite{Edwards1,Derro1,Maki1}.

Individual calculations of the chain and the plane reproduce well the
experimental spectra. This may suggest that the coupling between the
CuO chain and the neighboring CuO$_2$ planes is rather weak.

In summary, we have performed a RIXS study for the optimally doped
YBa$_2$Cu$_3$O$_{7-\delta}$, and found two characteristic excitations.
The excitation from the CuO$_2$ chain shows broad feature at 1.5-4 eV
and is almost independent of the momentum transfer, which is
consistent with the theoretical prediction. On the other hand, the
excitation from the CuO chain is enhanced at 2 eV near the zone
boundary of the $\bm{b}^{\ast}$ direction. The result indicates that
the Mott gap in the chain is smaller than that in the plane, and
provides an important constraint to understand how the CuO chain plays
a role in the superconductivity of YBCO.

K. T., T. T., and S.M, were supported by NAREGI Nanoscience Project
and Grantin-Aid for Scientific Research from the Ministry of
Education, Culture, Sports, Science and Technology of Japan.  The
numerical calculations were performed in the supercomputing facilities
in ISSP, Univ. of Tokyo and IMR, Tohoku Univ.. The crystal growth was
supported by the New Energy and Industrial Technology Development
Organization (NEDO) as the Collaborative Research and Development of
Fundamental Technologies for Superconductivity Applications.


\end{document}